\documentclass[conference]{IEEEtran}
\IEEEoverridecommandlockouts
\usepackage{color,soul}
\usepackage{cancel}
\usepackage{amsmath,amssymb,amsfonts, bm}
\usepackage{algorithmic}
\usepackage{graphicx} 
\usepackage{textcomp}
\def\BibTeX{{\rm B\kern-.05em{\sc i\kern-.025em b}\kern-.08em
    T\kern-.1667em\lower.7ex\hbox{E}\kern-.125emX}}

\usepackage[ruled,vlined]{algorithm2e} 
\usepackage{enumitem}
\usepackage[bottom]{footmisc}
\usepackage{color, xcolor, soul, framed}
\usepackage{cancel}
\usepackage[numbers,sort&compress]{natbib}
\usepackage[font=normalfont]{caption} 

\usepackage{subcaption}
\usepackage{booktabs}
\usepackage{mathtools}
\usepackage{graphicx}
\usepackage{hyperref}
\usepackage{varwidth}

\usepackage{multirow}
\usepackage{array}
\usepackage{lipsum}

\newcommand*{\vertbar}{\rule[-1ex]{0.5pt}{2.5ex}}

\newcommand{\TT}{\operatorname{T}}

\hypersetup{
    colorlinks=true,    citecolor=blue,    linkcolor=red,    filecolor=magenta,  urlcolor=blue,    pdftitle={Overleaf Example},
    }
    
\begin{document}

\title{Underwater MEMS Gyrocompassing:\\ A Virtual Testing Ground}

\author{Daniel~Engelsman,~\IEEEmembership{Graduate Student Member,~IEEE}, and Itzik Klein,~\IEEEmembership{Senior Member,~IEEE} 
\thanks{The authors are with the Charney School of Marine Sciences, Hatter Department of Marine Technologies, University of Haifa, Haifa, Israel. 

\noindent E-mails: \{dengelsm@campus, kitzik@univ\}.haifa.ac.il}}
\maketitle

\begin{abstract}
In underwater navigation, accurate heading information is crucial for accurately and continuously tracking trajectories, especially during extended missions beneath the waves. In order to determine the initial heading, a gyrocompassing procedure must be employed. As unmanned underwater vehicles (UUV) are susceptible to ocean currents and other disturbances, the model-based gyrocompassing procedure may experience degraded performance. To cope with such situations, this paper introduces a dedicated learning framework aimed at mitigating environmental effects and offering precise underwater gyrocompassing. Through the analysis of the dynamic UUV signature obtained from inertial measurements, our proposed framework learns to refine disturbed signals, enabling a focused examination of the earth's rotation rate vector. Leveraging recent machine learning advancements, empirical simulations assess the framework’s adaptability to challenging underwater conditions. Ultimately, its contribution lies in providing a resilient gyrocompassing solution for UUVs.
\end{abstract}

\section{Introduction}
Expressing the rotation angles about the principal axes of a three-dimensional coordinate frame is conveniently achieved using the Euler angles formalism \cite{titterton2004strapdown, farrell2008aided}. Determining the initial values of these angles, known as roll, pitch, and yaw, is crucial for effective navigation.
Similar to land and aerial vehicles, most marine platforms are designed to operate horizontally (leveled), maintaining their longitudinal axis aligned parallel to the local navigation plane most of the time.
To ascertain such spatial relationship between the x-y body plane and the north-East plane, accelerometers are employed to sense the local gravity. Subsequently, the roll and pitch angles can be computed from the axial components using analytical closed-form relations.
Utilizing modern micro-electro-mechanical systems (MEMS) sensors, successful leveling with an error of no more than 2 mrad is achievable, thanks to the pronounced gravity signal ensuring a high signal-to-noise ratio (SNR)  \cite{britting2010inertial}.
\\
Yet, in the process of orienting the longitudinal axis with respect to true north, i.e. finding the heading angle, most kinematic mechanisms heavily rely on external sources to provide the necessary platform information. For example, consumer-grade magnetometers, priced at just a few US dollars, prove efficient for tasks such as indoor orientation, pedestrian orientation, and low-fidelity navigation tasks.
\\
Other strategies include satellite systems, radio frequency (RF)-based navigation, and, more recently, fusion of visual odometry with the inertial navigation system (INS) solution.
%
While these methods prove practical for a wide variety of problems above sea level \cite{chatfield1997fundamentals}, they fail to operate or provide accurate initial heading while underwater. 
\\
Hence, earth's rotation rate ($\boldsymbol{\omega}^e_{ie}$) is employed to compute the carrier's initial heading angle in a process known as gyrocompassing. Relying solely on the gyroscopes sensitivity, this method obviate the need in external broadcasting and magnetic interference. However, the slow 24 hours rotation cycle generates an exceedingly subtle signal of 73$\textsc{e}$-6\,[rad/sec], detectable by only a few high-end sensors \cite{park1995covariance, renkoski2008effect}. 
%
With the advancement of MEMS technology, off-the-shelf gyroscopes have reached technological maturity, exhibiting sensitivity to noise levels as low as $\omega_{ie}$. Furthermore, recent machine learning and deep learning approaches have further advanced performance in various navigation tasks \cite{klein2022data, cohen2023inertial, cohen2024kit}.
\\
So while conventional noise removal methods restrict their operation to stationary conditions, this paper presents a realistic proof of concept of underwater gyrocompassing subjected to dynamic effects. From sensor input to model output, we analyze environmental disturbances and examine our model's capability to mitigate them.

\section{Problem Formulation}
This section lays the essential groundwork for understanding gyrocompassing theory, hypothesized UUV dynamics, and how they ultimately manifest through the inertial sensors.

\subsection{Gyrocompassing}
\begin{figure}[h] 
\begin{center}
\includegraphics[width=0.26\textwidth]{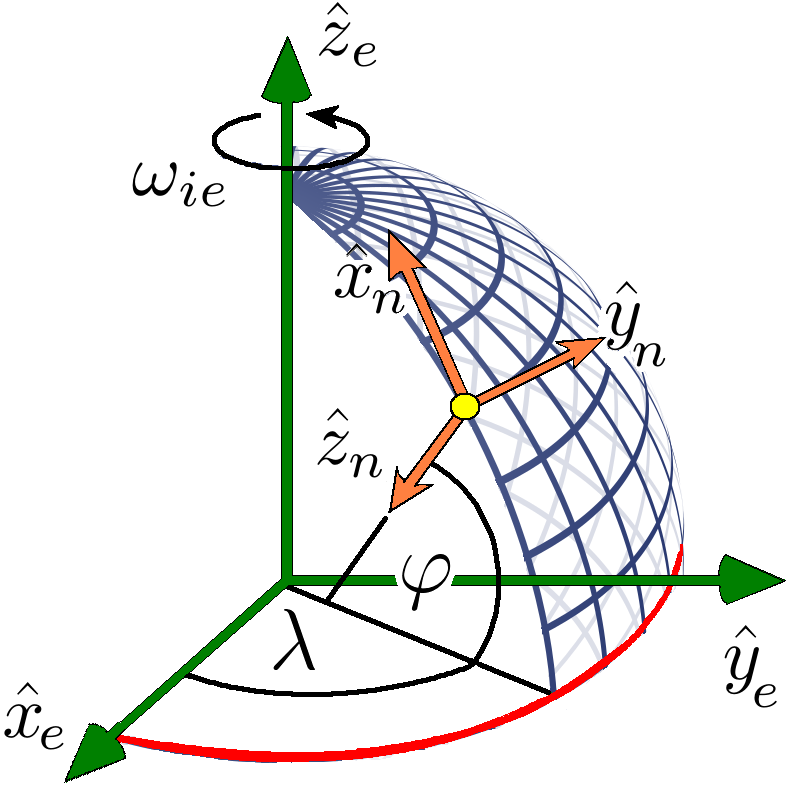}
\caption{Diagrammatic representation of the transformation between ECEF and navigation reference frames, using the latitude, $\phi$, and longitude, $\lambda$ angles.}
\label{fig:frame_1}
\end{center}
\end{figure}

Finding the north direction can be achieved through two distinct methods: Magnetic north, pointing towards the north magnetic pole, and true Geographic north, pointing towards earth's true north pole, around which earth revolves. Situated far apart, each pair of poles induces distinct rotational axes that do not coincide, thus resulting in magnetic declination. 
\\
Under reasonably sterile conditions, magnetic-based compassing provides a fair heading estimate while maintaining simplicity. However, when it comes to the complex UUVs design, their onboard electronics make them susceptible to electromagnetic interference, rendering magnetic-based compassing unreliable for underwater tasks. 
\\
In contrast, gyro-based navigation offers a standalone solution that remains robust across a wide variety of UUV tasks, a point further discussed in this article.
To begin with, rigid body orientation is parameterized through an Euler angle triad, utilizing an orthonormal transformation matrix that satisfies $\textbf{T}_i^b = (\textbf{T}_b^i)^{\TT}$, and $\| \textbf{T}_i^b \| = 1$. 
\\
By specifying the required rotations between a given body frame ($b$) and a local inertial frame ($i$), it involves three successive rotations—roll, pitch, and yaw ($\phi, \theta, \psi$)—about the x-y-z axes, as expressed by:
\begin{equation} \label{eq:T_b_n}
\textbf{T}_{i}^b = \begin{bmatrix}
\text{c}_\theta  \text{c}_\psi  &  \text{s}_\phi \text{s}_\theta \text{c}_\psi - \text{c}_\phi \text{s}_\psi  &   \text{c}_\phi \text{s}_\theta \text{c}_\psi + \text{s}_\phi \text{s}_\psi \\
\text{c}_\theta \text{s}_\psi & \text{s}_\phi \text{s}_\theta \text{s}_\psi +  \text{c}_\phi \text{c}_\psi & \text{c}_\phi \text{s}_\theta \text{s}_\psi - \text{s}_\phi \text{c}_\psi  \\
-\text{s}_\theta & \text{s}_\phi \text{c}_\theta & \text{c}_\phi \text{c}_\theta 
\end{bmatrix} \ ,
\end{equation}
\\
with 's', and 'c', represent the sine and cosine functions, respectively. We define $\boldsymbol{\omega}_{ie}^e$ as the earth rotation vector, where $e$ denotes the earth centered earth-fixed (ECEF) frame. Fig.~\ref{fig:frame_1} illustrates the rotation of $\boldsymbol{\omega}_{ie}^e$ about the z-axis of the ECEF frame and its projection onto the body frame by
\begin{align} \label{eq:omega_ie_e}
\boldsymbol{\omega}_{ie}^b = \textbf{T}_{n}^b \textbf{T}_{e}^n
\boldsymbol{\omega}_{ie}^e \ ,
\end{align}
where $\textbf{T}_{e}^n$ represents earth to navigation frame transformation, followed by $\textbf{T}_{n}^b$, which signifies the navigation to body frame transformation. For simplicity, and without the lose of generality, we assume that the gyroscopes sensitive axes align with the body axes such that the gyro output is:
\begin{align} \label{eq:omega_ie_b}
\boldsymbol{\omega}_{ib}^b = \begin{bmatrix} \ p & q & r \ \ \end{bmatrix}^{\TT} \ .
\end{align}
Thus, in stationary conditions when combining \eqref{eq:omega_ie_e} and \eqref{eq:omega_ie_b}, the heading angle is confined through the following closed-form gyrocompassing expression \cite{groves2015principles}
\begin{align}
\text{s}_{\psi} &= -q \, \text{c}_{\phi} + r \, \text{s}_{\phi} \ , \\
\text{c}_{\psi} &= \ p \, \text{c}_{\theta} + q \, \text{s}_{\phi} \text{s}_{\theta} + r \, \text{c}_{\phi} \text{s}_{\theta} \ ,
\end{align}
and can be solved with a four-quadrant arctangent function by
\begin{align} \label{eq:psi_GC}
\underline{\psi} (p, q, r) &= \arctan_2 \big( \text{s}_{\psi} \, , \, \text{c}_{\psi} \big) \ . 
\end{align}
Given proper leveling, or equivalently, under horizontal conditions ($\phi=\theta=0$), \eqref{eq:psi_GC} simplifies into a more streamlined expression
\begin{align} \label{eq:psi_GC_simp}
\underline{\psi} (p, q) = \arctan_2 \big( -q \, , \, p \big) \ ,
\end{align}
revealing that the predominant component of $\boldsymbol{\omega}_{ie}^e$ lies only within the gyroscope x-y plane.

\subsection{Gyroscope Measurement Model} \label{s:gyro_model}
In general, instrumental errors are typically categorized into \textit{stochastic} and \textit{deterministic} sources. The former is unmodelable due to an unknown origin, fluctuates randomly between time steps, and maintains a constant variance upon remeasurement. The latter exhibits constancy between time steps but may vary run-to-run. 
Conventional linearization schemes establish a relationship between the gyro outputs $\boldsymbol{\tilde{\omega}}_{b}$ and the ground truth (GT) quantities $\boldsymbol{\omega}_{b}$, characterized by
\begin{align} \label{eq:system}
\boldsymbol{\tilde{\omega}}_{b} = \boldsymbol{M} \boldsymbol{\omega}_{b} + \textbf{\textit{b}}_g + \textit{\textbf{w}}_g \ .
\end{align}
Here, $\boldsymbol{\tilde{\omega}}_{b}$ is directly influenced by a scale-factor and misalignment matrix $\boldsymbol{M}$, affecting its intensity. Concurrently, $\textbf{\textit{b}}_g$ represents the uncompensated bias, and $\textit{\textbf{w}}_g$ introduces additional signal obscuration, modeled as a zero-mean white Gaussian noise \cite{woodman2007introduction, engelsman2023parametric}. Evidently, deterministic errors manifest as linear mappings from noisy inputs to their respective outputs.

\subsection{Rigid Body Kinematics}
\begin{figure}[h]
\begin{center}
\includegraphics[width=0.46\textwidth]{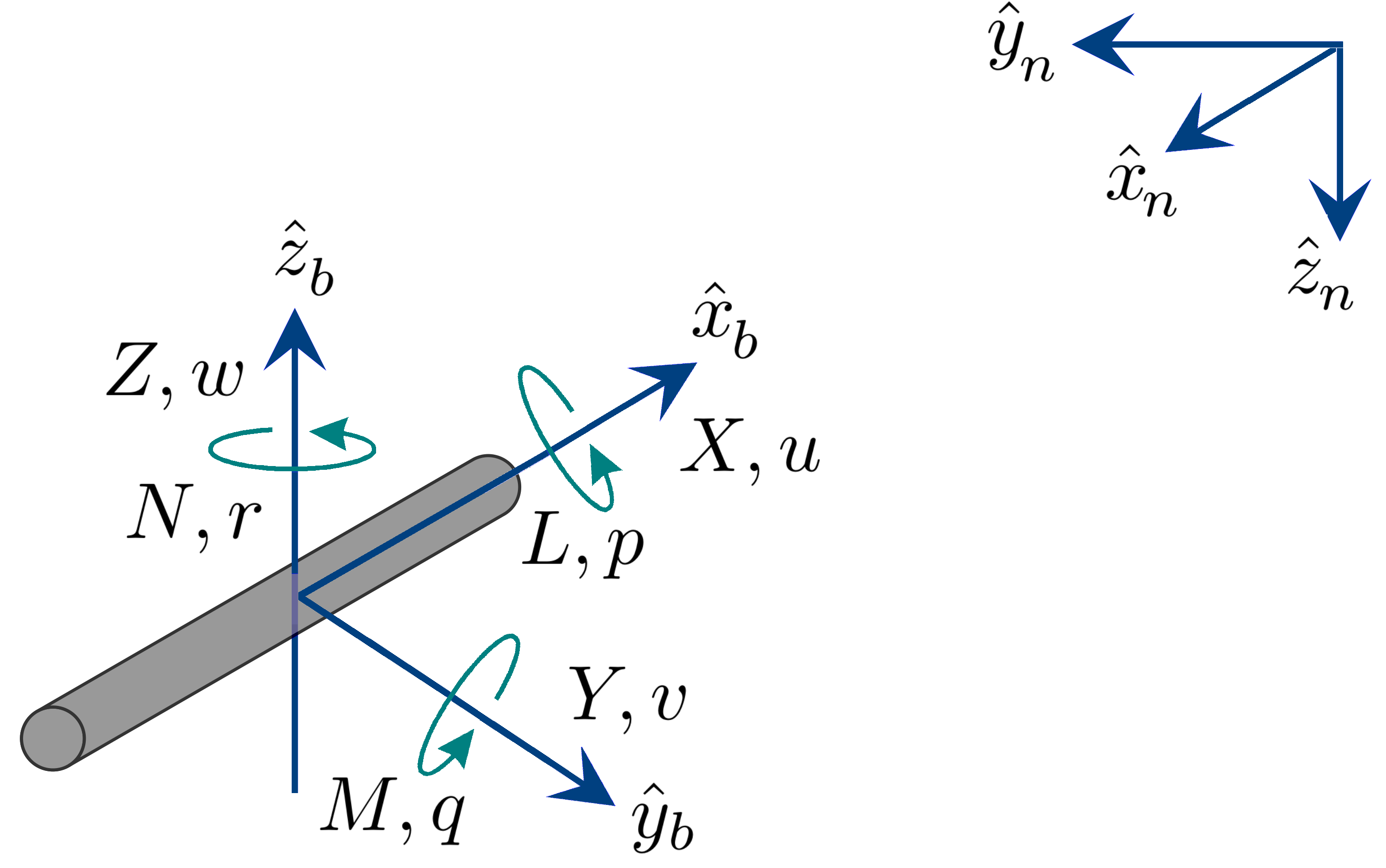}
\caption{Relationship between the navigation frame and UUV body reference frames.}
\label{fig:frame_2}
\end{center}
\end{figure}

A six degrees of freedom (DoF) motion model is defined by one set of 3D coordinates addressing linear translations and another set addressing the associated rotations.
Let the following state vector denote the position and orientation in the inertial frame 
\begin{align} \label{eq:vec-inertial}
\boldsymbol{\eta}_{b} = \begin{bmatrix} \ \boldsymbol{\eta}_1^{\TT} \ | \ \boldsymbol{\eta}_2^{\TT} \ \  \end{bmatrix}^{\TT} = \begin{bmatrix} \ x & y & z \ \  |  \ \ \phi & \theta & \psi \ \end{bmatrix}^{\TT} \, ,
\end{align}
and similarly, the 6-DoF body velocities are given by
\begin{align} \label{eq:vec-body}
\boldsymbol{\nu}_i = \begin{bmatrix} \ \boldsymbol{\nu}_1^{\TT} \ | \ \boldsymbol{\nu}_2^{\TT} \ \  \end{bmatrix}^{\TT} = \begin{bmatrix} \ u & v & w \ \  |  \ \ p & q & r \ \, \end{bmatrix}^{\TT} \, .
\end{align}
\begin{table}[b]
\renewcommand{\arraystretch}{1.45}
\begin{center}
\begin{tabular}{c|c|c|c||c|c|} \cline{2-6}
 & \shortstack{{}\\DoF\\{}} & \shortstack{Forces \&\\ \ moments} & \shortstack{{}\\Notation\\{}} & \shortstack{Inertial\\ \ frame} & \shortstack{{}\\Body \\ \hspace{1mm} frame} \\ \specialrule{1.15pt}{1pt}{1pt}
\multicolumn{1}{|c|}{ \multirow{3}{*}{Translational} } & 1 & Heave & $X$ & x & $u$ \\ \cline{2-6}
\multicolumn{1}{ |c| }{}& 2 & Sway & $Y$ & y & $v$ \\ \cline{2-6}
\multicolumn{1}{ |c| }{}& 3 & Surge & $Z$ & z & $w$ \\ \specialrule{1.15pt}{1pt}{1pt}
\multicolumn{1}{ |c| }{ \multirow{3}{*}{Rotational} } & 4 & Rolling & $L$ & $\phi$ & $p$ \\ \cline{2-6}
\multicolumn{1}{ |c| }{}& 5 & Pitching & $M$ & $\theta$ & $q$ \\ \cline{2-6}
\multicolumn{1}{ |c| }{}& 6 & Yawing & $N$ & $\psi$ & $r$ \\ \hline
\end{tabular}
\end{center}
\caption{Key notations in UUV motion variables.} \label{t:params}
\end{table}
Unlike the orthonormal transformation \eqref{eq:T_b_n} which relates translational velocities between inertial and body frames, the mapping of body-axis rates to Euler angle rates ($\dot{\phi}, \dot{\theta}, \dot{\psi}$) is carried out through
\begin{equation} \label{eq:T_b_n_dot}
\textbf{T}_{\dot\Omega}^{ \, b} = \begin{bmatrix}
1 & \text{s}_\phi \text{tan}_\theta & \text{c}_\phi \text{tan}_\theta \\
0 & \text{c}_\phi & -\text{s}_\phi \\
0 & \text{s}_\phi / \text{c}_\theta & \text{c}_\phi / \text{c}_\theta 
\end{bmatrix} \ .
\end{equation}
Therefore, by jointly combining \eqref{eq:T_b_n} and \eqref{eq:T_b_n_dot}) into 
\begin{align}
\textbf{T} = \begin{bmatrix} \textbf{T}_{i}^b & 
 \boldsymbol{0}_{3 \times 3} \ \ \\
 \ \boldsymbol{0}_{3 \times 3} &  \textbf{T}_{\dot\Omega}^{ \, b} \end{bmatrix} \ ,
\end{align}
the kinematic model is given in vectorial representation by
\begin{align} \label{eq:eta_dot}
\boldsymbol{\dot{\eta}} = \textbf{T} \, \boldsymbol{{\nu}} \ .
\end{align}
%
%
Table~\ref{t:params} compiles the relevant nomenclature, assuming that the sensor frame is aligned with the UUV center of mass. This arrangement expects the accelerometers to sense the induced forces ($X,Y,Z$) along the longitudinal ($\hat{{x}}_b$), transverse ($\hat{{y}}_b$), and vertical axes ($\hat{{z}}_b$), while the gyros will detect the angular velocities resulting from the 3D moments ($L,M,N$) around these axes.
%
To visualize this, Fig.~\ref{fig:frame_2} provides simplified schematics of the dynamics acting along the UUV body axes and the local navigation frame.
\\
Despite extensive literature coverage on control aspects, this work focuses solely on observing vehicle states under external interruptions, such as characteristic underwater forces and moments \cite{nahon1996simplified, prestero2001verification, gomes2005underwater}. Hence, control inputs are disregarded to maintain focus solely on gyrocompassing.

\subsection{Rigid Body Dynamics}
While \textit{kinematics} studies the object's motion without considering the driving causes, \textit{dynamics}, on the other hand, explores the cause-and-effect relationships underlying the observed motion. To begin with, Newton's second law states that the net force exerted on an object of mass ($m$) is directly proportional to its linear acceleration ($\dot{\boldsymbol{\nu}}_1$), as given in
\begin{align} \label{eq:forces}
\boldsymbol{\tau}_1 = m \ \dot{\boldsymbol{\nu}}_1 \ .
\end{align}
The rotational analog is provided by Euler's second law, which relates changes in angular momentum to external torques \cite{fossen1999guidance}
\begin{align} \label{eq:moments}
\boldsymbol{\tau}_2 = \boldsymbol{I}_0 \, \dot{\boldsymbol{\omega}}_b + \boldsymbol{\omega}_b \times (\boldsymbol{I}_0 \, \boldsymbol{\omega}_b) \approx \boldsymbol{I}_0 \, \dot{\boldsymbol{\nu}}_2 \ ,
\end{align}
where $\boldsymbol{I}_0$ represents a 3D inertia tensor, and $\boldsymbol{I}_{3}$ is a $3\times3$ identity matrix. After augmentation with \eqref{eq:forces}, the external factors acting on the UUV are collectively determined by
\begin{align}
\boldsymbol{\tau} = \begin{bmatrix} \ \boldsymbol{\tau}_1^{\TT} \ | \ \boldsymbol{\tau}_2^{\TT} \ \ \end{bmatrix}^{\TT} = 
\begin{bmatrix} \ X & Y & Z \ \ | \ \ L & M & N \ \ \end{bmatrix}^{\TT} ,
\end{align}
resulting in the following 6-DoF rigid body model:
\begin{align} \label{eq:6dof-full}
\boldsymbol{M} \, \dot{\boldsymbol{\nu}} + \boldsymbol{C}(\boldsymbol{\nu}) \, \boldsymbol{\nu} + \boldsymbol{D}(\boldsymbol{\nu}) \,  \boldsymbol{\nu} + \boldsymbol{g}(\boldsymbol{\eta}) = \boldsymbol{{\tau}} \ .
\end{align}
The mass-inertia matrix, $\boldsymbol{M}$, maps the state vector $\dot{\boldsymbol{\nu}}$ to the associated mass and inertia terms (Appendix \ref{appendix:a}), and the skew-symmetric operator is given by $[ \, \cdot \, ]_{\times}:\mathbb{R}^3 \rightarrow \mathfrak{so}(3)$. 
\\
In a bottom-heavy configuration, the center of gravity is positioned below the body-frame origin, i.e. $\boldsymbol{r}_W = \begin{bmatrix}
0 & 0 & z_W \end{bmatrix}^{\TT}$, yielding the following $6 \times 6$ structure
\begin{align}
\boldsymbol{M} = \begin{bmatrix}
m \, \boldsymbol{I}_{3} & -m \, [\boldsymbol{r}_W]_{\times} \\
m \, [\boldsymbol{r}_W]_{\times} & \boldsymbol{I}_0
\end{bmatrix} \ .
\end{align}
At the water-air interface, surface vessels experience frequent wave disturbances—a dynamic that involves cross-coupling between $\boldsymbol{\nu}$ and $\dot{\boldsymbol{\nu}}$—expressed in the Coriolis matrix $\boldsymbol{C}$. In contrast, subsea vehicles are exposed to smaller amplitudes and slower oscillations induced by ocean currents, leading to an approximation of $ \boldsymbol{C} \approx 0$. This facilitates a critical underwater capability known as $ \textit{station keeping} $—a disturbance-free period that enables a stable hovering while executing intricate tasks such as localization and mapping.
\\
The third term in \eqref{eq:6dof-full} is the damping matrix ($\boldsymbol{D}$), which captures the body's resistive forces countering its motion in the medium. 
Assuming uncoupled motion, its approximate coefficients suggest the following diagonal structure
\begin{align}
\boldsymbol{D} ({\boldsymbol{\nu}}) = - \operatorname{diag} \{ \, X_u, \, Y_v, \, Z_w, \, L_p, \, M_q, \, N_r \, \} \ .
\end{align}
The concluding term, $\boldsymbol{g}(\boldsymbol{\eta})$, signifies the restoring forces and moments that actively counteract disruptions in the delicate weight ($\boldsymbol{W}$) buoyancy ($\boldsymbol{B}$) equilibrium \cite{newman2018marine}, formulated as
\begin{align}
\boldsymbol{f}_W ( \boldsymbol{\eta}_2 ) &= +\textbf{T}_{b}^i \, ( \boldsymbol{\eta}_2 ) \, 
(\boldsymbol{W} \cdot \hat{z}_b) \ , \\
\boldsymbol{f}_B ( \boldsymbol{\eta}_2 ) &= -\textbf{T}_{b}^i \, ( \boldsymbol{\eta}_2 ) \, (\boldsymbol{B} \, \cdot \hat{z}_b) \ .
\end{align}
Maintaining a stable upright position depends on the body's resistance to externally induced reactive moments. With the center of buoyancy positioned in the upper $xy$ half-plane, $\boldsymbol{r}_B = \begin{bmatrix} 0 & 0 & z_B \end{bmatrix}^{\mathrm{T}}$, the restoring force vector can be written in a compact form as
\begin{align}
\boldsymbol{g}(\boldsymbol{\eta}) = - \begin{bmatrix}
\boldsymbol{f}_W ( \boldsymbol{\eta}_2 ) + \boldsymbol{f}_B ( \boldsymbol{\eta}_2 ) \\
\boldsymbol{r}_W \times \boldsymbol{f}_W ( \boldsymbol{\eta}_2 ) + \boldsymbol{r}_B \times \boldsymbol{f}_B ( \boldsymbol{\eta}_2 ) \end{bmatrix} \ .
\end{align}
Ultimately, our Newton-Euler equations in body coordinates take the form of
\begin{align} \label{eq:6dof-final}
\boldsymbol{M} \, \dot{\boldsymbol{\nu}} + \boldsymbol{D}(\boldsymbol{\nu}) \,  \boldsymbol{\nu} + \boldsymbol{g}(\boldsymbol{\eta}) = \boldsymbol{{\tau}} \ .
\end{align}
For an inertial-fixed frame, $\boldsymbol{{\nu}}$ can be replaced by inverting \eqref{eq:eta_dot} and taking its time derivative approximation, namely
\begin{align} \label{eq:substitution}
\boldsymbol{{\nu}} =& \ \textbf{T}^{-1} \, \boldsymbol{\dot{\eta}} \ , \\
\boldsymbol{\dot{{\nu}}} =& \ \cancelto{0}{\dot{\textbf{T}  }^{-1}} \boldsymbol{\dot{\eta}} + \ {\textbf{T}}^{-1} \, \boldsymbol{\ddot{\eta}} \approx \ {\textbf{T}}^{-1} \, \boldsymbol{\ddot{\eta}} \ .
\end{align}

\section{Proposed Approach} \label{s:methodology}
This section introduces our approach to combining the established theory for understanding both gyros errors and simplified UUV dynamics, aiming to investigate the feasibility of underwater MEMS gyrocompassing.

\subsection{The Big Picture}
\begin{figure}[h]
\begin{center}
\includegraphics[width=0.5\textwidth]{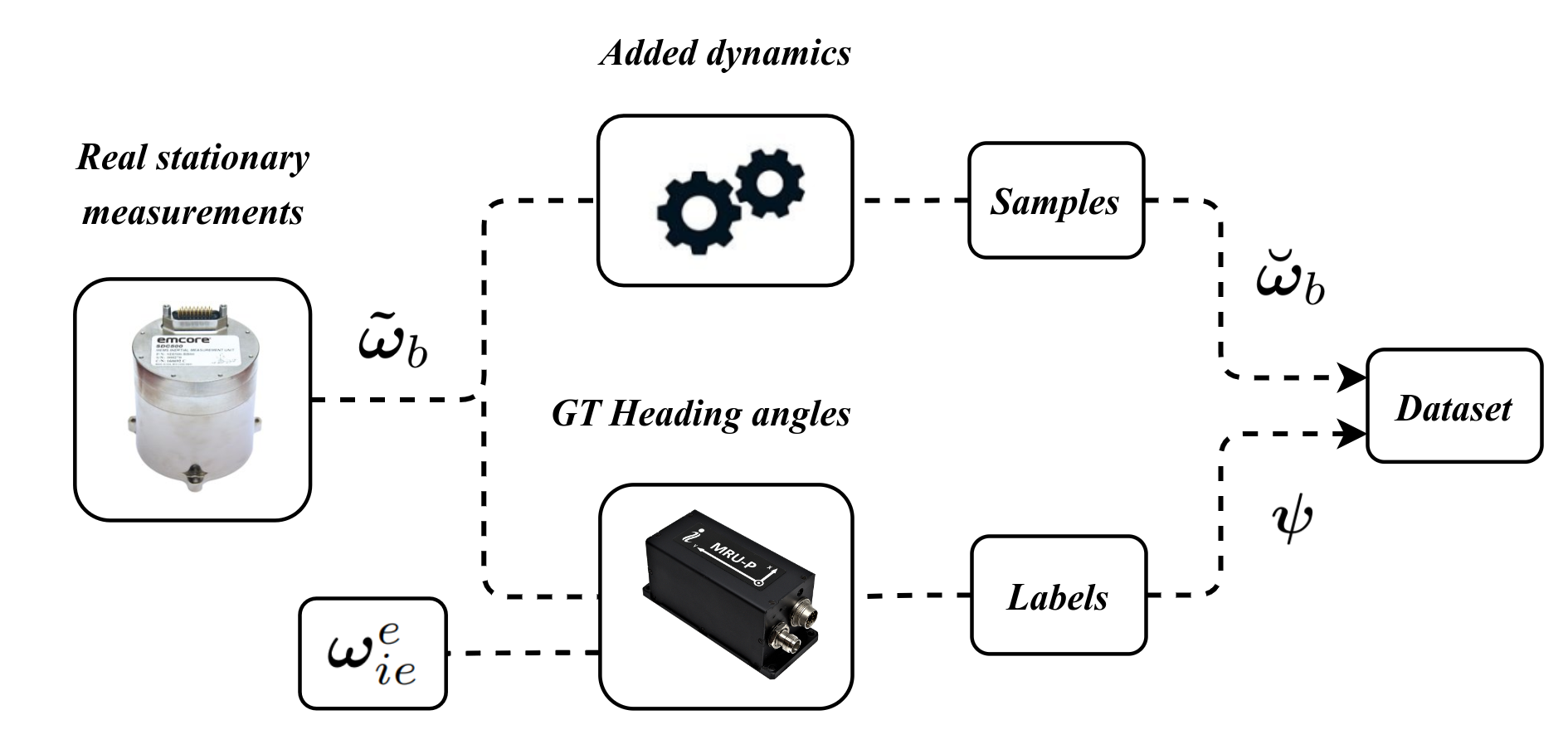}
\caption{Conceptual flow of the proposed setup.}
\label{fig:big_picture}
\end{center}
\end{figure}

In our earlier study \cite{engelsman2023towards}, a commercially available Emcore device\footnote{Sensor specs @ \href{https://emcore.com/products/sdn500-ins-gps-tactical-grade-navigation-system/}{https://Emcore.com/products/SDN500}} was employed to measure angular rates on a stationary unit that can be freely rotated while leveled. With the primary objective of discerning how $\boldsymbol{\omega}_{ie}$ is affected by heading angle variations, a deep learning (DL) model\footnote{Data \& code @ \href{https://github.com/ansfl/Learning-Based-MEMS-Gyrocompassing}{https://GitHub.com/ansfl/LB-Gyrocompassing}} was proposed as means of tackling the weak rotation signal, thus enabling a direct heading estimate.
%
Here, we propose a superposition ($\oplus$) between real-world stationary measurements ($\boldsymbol{\tilde{\omega}}_{b}$) and our self-generated dynamics ($\boldsymbol{\breve{\omega}}_{b}$), resulting in the following synthesis
\begin{align} \label{eq:superimpose}
\boldsymbol{\breve{\omega}}_{b} = \boldsymbol{\tilde{\omega}}_{b} \, \oplus \, \boldsymbol{\check{\omega}}_{b} \ .
\end{align}
Building upon the prior success of our DL model in learning from stationary samples, the analysis of synthetic dynamics can now provide valuable insights into the upper limits of environmental disturbances under which gyrocompassing can be sustained.
Fig.~\ref{fig:big_picture} conceptualizes this integration architecture, resulting in a synthesized trainable dataset.

\subsection{Simulated Dynamics}
The complete 6-DoF dynamic model \eqref{eq:6dof-final} establishes the analytical relationship between linear accelerations and angular velocities, correlating them with external forces and moments, respectively. Upon further examination of the gyrocompassing equation \eqref{eq:psi_GC_simp}, it becomes evident that accelerations ($\dot{\boldsymbol{\nu}}_1$) have no impact on it, rendering it translation-invariant. 
\\
Conversely, angular velocities  ($\dot{\boldsymbol{\nu}}_2$) do indeed affect the heading estimate, establishing its rotation-variance property. 
Mathematically, this eliminates the ${\boldsymbol{\nu}}_1$ equations, reducing the system to be solely subjected to external moments $\boldsymbol{{\tau}}_2$, which act as the forcing term in the right hand side of
\begin{align} \label{eq:simplified}
\boldsymbol{I}_0 \, \dot{\boldsymbol{\nu}}_2 + \boldsymbol{D}_{2}(\boldsymbol{\nu}_2) \,  \boldsymbol{\nu}_2 + \boldsymbol{g}_{2} (\boldsymbol{\eta}_2) = \boldsymbol{{\tau}}_2 (t) \ ,
\end{align}
where the subscript '$2$' denotes rotational coordinates. Various disturbance models are discussed in details in \cite{fossen1999guidance} (page 71), and \cite{xue1997three}. However, to ensure the practicality of our simulation scheme, we confine our analysis to three distinct modes: 
\begin{enumerate}[label=(\roman*)]
    \item Impulse $\boldsymbol{{\tau}}_2 (t) = \delta(t)$: Signifies an instantaneous perturbation at $t_0$, resulting in short-term system responses.
    \item Step $\boldsymbol{{\tau}}_2 (t) = \mathbf{1}_{[0, t_{ss})}(t)$: Marking a sudden change at $t_0$, characterized by a transient behavior, subsequently transitioning to a decay towards a steady-state.
    \item Sinusoidal $\boldsymbol{{\tau}}_2 (t) = \mathcal{W} \cos(\omega_0 t + \varphi_0)$: Featuring periodic wave-like patterns with amplitude $\mathcal{W}$ and phase $\varphi_0$. The resulting oscillatory response decays in proportion to $\boldsymbol{D}$.
    %
    %
\end{enumerate}
Upon normalizing \eqref{eq:simplified}, the system manifests as a set of linear ordinary differential equations (ODE), solvable either by integrating factors or the Laplace transform (Appendix \ref{appendix:b}). 
While the system parameters naturally remain constant, the parameters of $\boldsymbol{{\tau}}_2$ vary randomly between iterations. 
\\
This strategy enhances the model's generalizability by closely simulating underwater conditions, thereby bridging the reality gap during inference.

\subsection{A Learning-Based Approach}
Though the analytical structure presented thus far might suggest the possibility of a closed-form solution, two main reasons defer such an assumption. Firstly, the initial conditions, i.e., body Euler angles, are not known a priori, particularly at depths of tens of meters. Secondly, the presence of time-varying dynamics renders conventional tools such as the Fourier transform or wavelet analysis suboptimal. To that end, learning approaches excel in adapting and generalizing from data, especially in our case where the meaningful signal $\boldsymbol{\omega}_{ie}^e$, is 'buried' under both instrumental errors $\boldsymbol{\tilde{\omega}}_{b}$ and complex underwater dynamics $\boldsymbol{\check{\omega}}_{b}$. 
\\
Within a supervised learning paradigm, the problem formulation differentiates the input sample space \(\boldsymbol{\Omega}\) form the corresponding label space \(\boldsymbol{\psi}\), linked by a statistical probability measure \(\mathcal{P}\). A learner in the hypothesis space \(f \in \mathcal{F}\) is tasked with performing an inference mapping \(f: \boldsymbol{\Omega} \rightarrow \boldsymbol{\psi}\), aiming to minimize generalization error \(\mathcal{L}_{\mathcal{P}}\) as follows
\begin{align}
\boldsymbol{\theta}^* = \arg \min_{\boldsymbol{\theta}} \mathcal{L}_{\mathcal{P}} = \mathbb{E}_{(\omega, \psi)\sim \mathcal{P}} \left\{ \ell \left( \hat{\psi}, \psi \right) \right\} \ ,
\end{align}
where \(\boldsymbol{\theta}\) denotes the tuned parameters, \(\hat{\psi} = f_\theta (\boldsymbol{\breve{\omega}}_{b})\) expresses the model predictions (heading angles), and the inference is evaluated using quadratic loss 
\begin{align}
\ell : \mathcal{F} \times \boldsymbol{\Omega} \times \boldsymbol{\psi} \rightarrow \mathbb{R}_{+} \ .
\end{align}
\begin{algorithm}[b]
\caption{The augmented dynamics algorithm} 
\label{alg:aug}
\SetAlgoLined
\SetKwInOut{Input}{Input}
\SetKwInOut{Output}{Output}
\Input{Labeled sensor data ( $\boldsymbol{\tilde{\Omega}} , \boldsymbol{\psi}$ )}
\SetKwFunction{Tensor}{torch.tensor}
\SetKwFunction{FMain}{dataAugment}
{
    $\boldsymbol{\breve{\Omega}} \leftarrow$ [~] \tcp*{Initialization}
    \For{$(i \, , \, \boldsymbol{\tilde{\omega}}_i)$ in $\boldsymbol{\tilde{\Omega}}$}{
        $\boldsymbol{\tau}_2 (t) \leftarrow \texttt{generate}( \mathcal{W}, \, \omega_0, \, \varphi_0 )$ \\
        $\boldsymbol{\check{\omega}}_i \leftarrow \texttt{solve}( \ref{eq:simplified} ) $ \\
        $\boldsymbol{\breve{\omega}}_{i}  \leftarrow \boldsymbol{\tilde{\omega}}_{i} \, \oplus \, \boldsymbol{\check{\omega}}_{i}$ \\ 
        $\boldsymbol{\breve{\Omega}}\texttt{.add}$( $\boldsymbol{\breve{\omega}}_{i}$ ) \\
    }
}
\KwRet \texttt{tensor}( $\boldsymbol{\breve{\Omega}} , \boldsymbol{\psi}$ ) \\
\end{algorithm}
Algorithm~\ref{alg:aug} outlines the additive process by which the stationary dataset $\boldsymbol{\tilde{\Omega}}$ incorporates artificial dynamics $\boldsymbol{\check{\Omega}}$. 
\\
During the $i$-th iteration, simulation parameters are randomly drawn ($\texttt{generate}$) to allow for a wide range of excitation profiles $\boldsymbol{\tau}_2 (t)$. With all variables assigned, further substitution in the system equation facilitates the synthesis of disturbances ($\texttt{solve}$), resulting in a new mixed-mode volume $\boldsymbol{\breve{\Omega}}$.

\section{Empirical Assessment}
\begin{figure}[t]
\begin{center}
\includegraphics[width=0.47\textwidth]{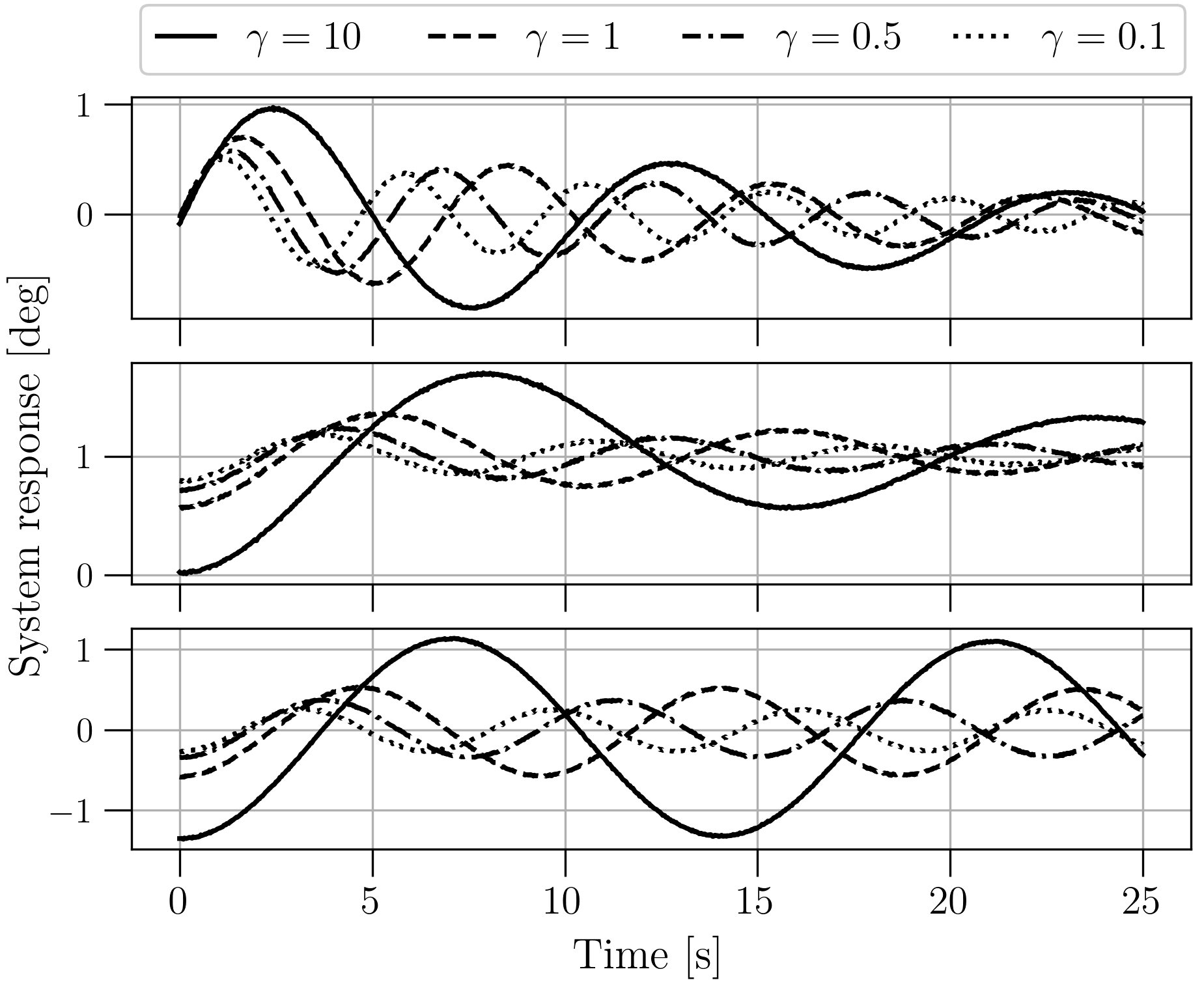}
\caption{Simulation of the characteristic excitation profiles: impulse (top), step (middle), and sinusoidal (bottom) functions.}
\label{fig:dynamics}
\end{center}
\end{figure}

\begin{figure}[t]
\begin{center}
\includegraphics[width=0.48\textwidth]{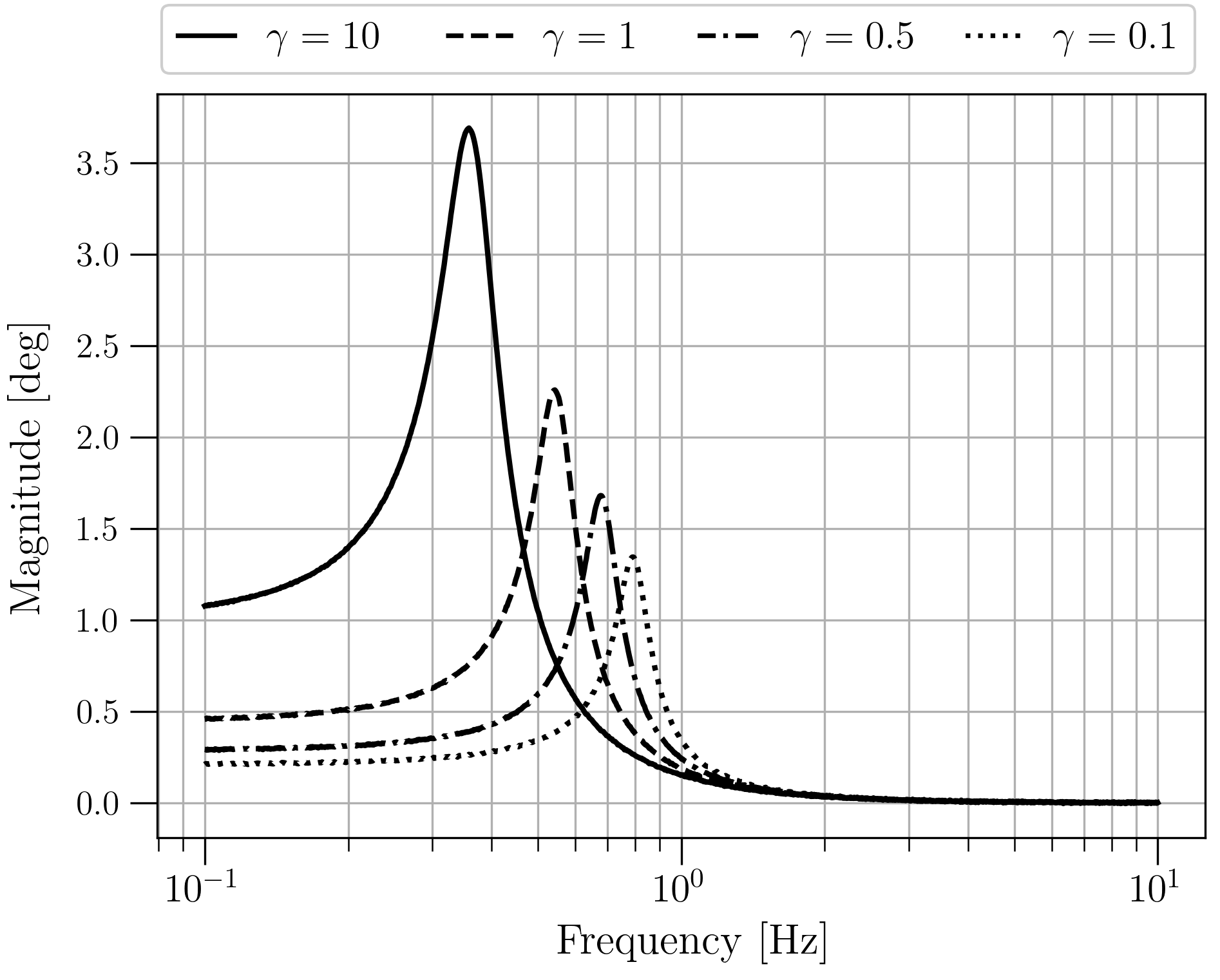}
\caption{Spectral response to unit input $\boldsymbol{\tau}(t)$.}
\label{fig:spectra}
\end{center}
\end{figure}

\begin{figure}[t]
\begin{center}
\includegraphics[width=0.48\textwidth]{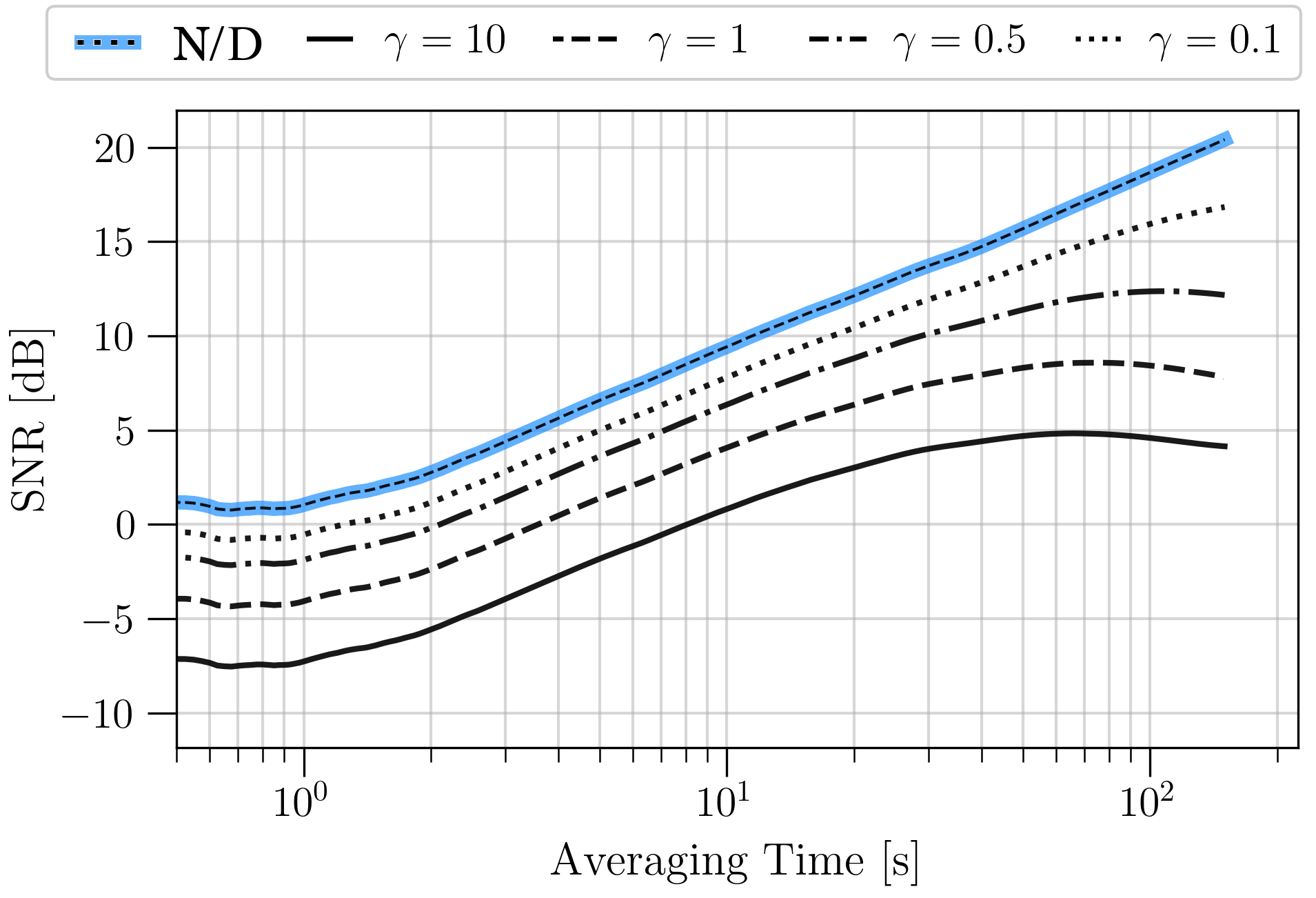}
\caption{SNR comparison with varying averaging time and $\gamma$'s.}
\label{fig:SNR}
\end{center}
\end{figure}

\begin{figure}[h]
\begin{center}
\includegraphics[width=0.48\textwidth]{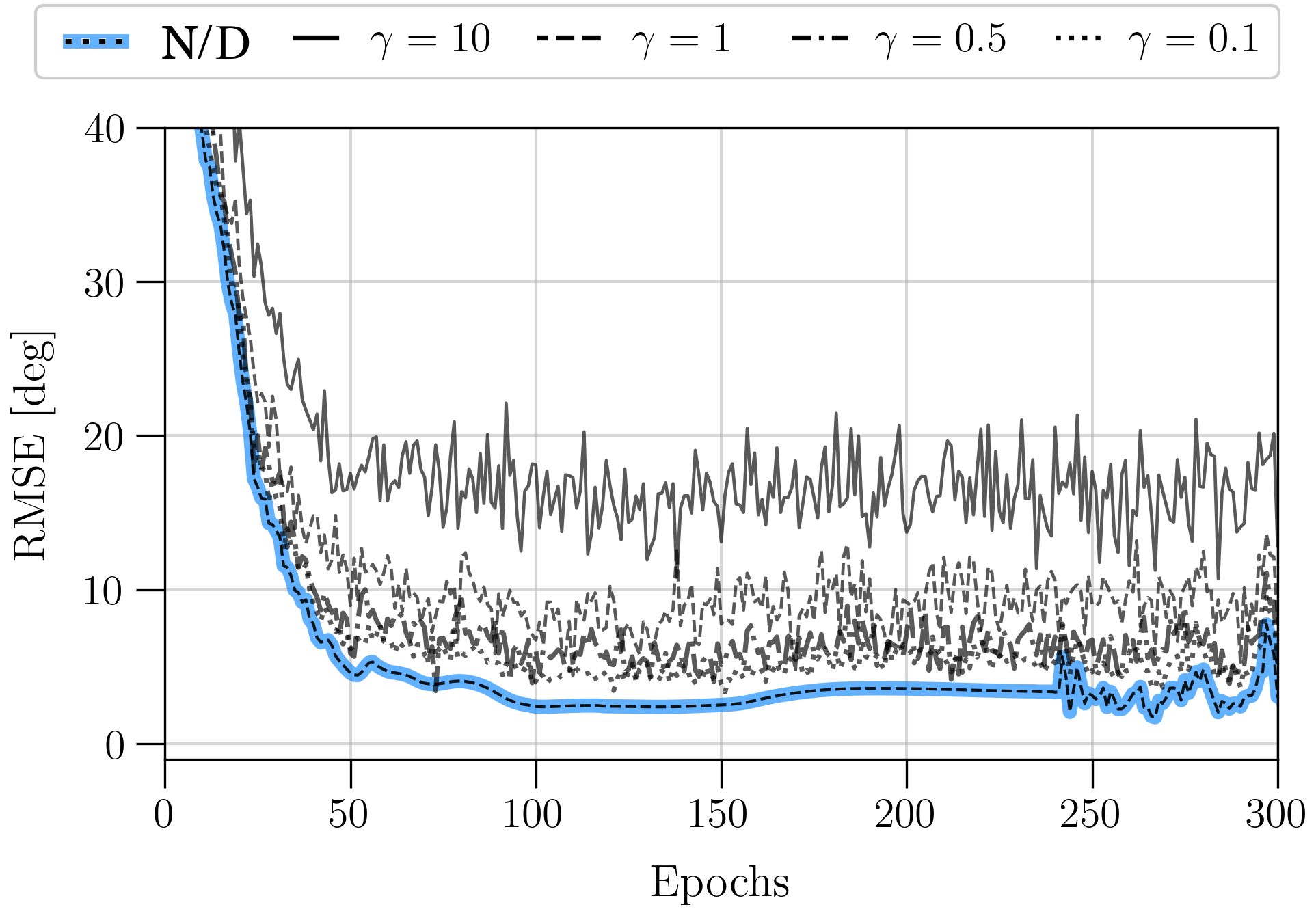}
\caption{Validation curve against number of epochs.}
\label{fig:loss}
\end{center}
\end{figure}

Assuming the intrinsic properties $\boldsymbol{I}_0$ and $\boldsymbol{D}_{2}$ to be time-invariant, the system response seems to be dominated by the torque-to-inertia ratio $\left( \frac{ \boldsymbol{\tau} (t) }{ \boldsymbol{I}_{0} } \right)$, or $\gamma$ for brevity.
\\
This normalization term ensures geometry-independence, allowing a focus on the order of magnitude (OoM) regardless of specific configurations. Furthermore, it facilitates the integration of varying disturbance intensities, enhancing the model's ability to learn and adapt to characteristic responses.
\\
Figure~\ref{fig:dynamics} illustrates the UUV's dynamic response, showcasing three distinct solutions derived from \eqref{eq:simplified} and unveiling the system's angular response to diverse inputs. When subjected to a wave input, the UUV exhibits an initial angle tilt, revealing an oscillatory relationship intricately linked to its specific geometry and mass properties.
The top two plots capture the system's reaction to impulse and step inputs, demonstrating transient behavior originating from the damping effect. Observably, the angular states exhibit a gradual relaxation process before reaching a steady-state. On the contrary, the bottom plot illustrates the system's response to a sinusoidal input, underscoring the preservation of input periodicity. Notably, variations in scaling and phase shift in the sinusoidal input are attributed to the inherent properties of the system.
\\
Subsequently, a detailed analysis of $\boldsymbol{\eta}_2$ (see \eqref{eq:oscillatory} in Appendix \ref{appendix:b}) is conducted to discern the spectral response of the system, shedding light on the frequencies experiencing amplification and attenuation. In marine environments, ocean currents typically manifest at relatively low frequencies. 
\\
Fig.~\ref{fig:spectra} investigates the spectral content of the angular response to characteristic inputs. Ocean currents manifest a broad spectrum of frequencies, ranging from the gradual rhythms of slow tidal waves to the more frequent patterns associated with stormy seas. In this context, the damping effect is notably pronounced at higher frequencies, wherein the opposing force of water resistance effectively counters oscillations, resulting in a gradual attenuation.
\\
As observed, lower values of $\gamma$ correspond to decreased amplitudes in the system's response. This is a predictable outcome, as less wave energy is considered in such scenarios. Nevertheless, these lower values provide valuable insights into the interplay between the UUV and its surrounding environment.
Finally, to comprehend how everything converges in the gyrocompassing procedure, let us examine the dynamics' influence on signal clarity in terms of SNR, given by 
\begin{align}
\text{SNR}_{\text{dB}} = \frac{P_{\text{signal,dB}}}{P_{\text{noise,dB}}} = 20 \log_{10} \left( \frac{A_{\text{signal}}}{A_{\text{noise}}}  \right) \ ,
\end{align}
where $P$ stands for power and $A$ stands for amplitude, and expressed in decibels (dB). To visualize that, Fig.~\ref{fig:SNR} depicts the gyroscopes SNR versus averaging time. The computation of their amplitudes involves dividing the square of $\boldsymbol{\omega}_{ie}^e$ by the square norm of the gyros' standard deviation. In addition to the four $\gamma$ values observed previously, the thick blue line represents a non-dynamic scenario (N/D), effectively serving as a control group. Being stationary, it exhibits an expected linear growth of SNR by 20 decibels (1 OoM) per decade, on a logarithmic time scale. 
\\
The challenging task of signal enhancement becomes evident: the stronger the waves impacting the UUV, the more undesirable frequencies propagate within the signal, compromising $\boldsymbol{\omega}_{ie}^e$ clarity, and thereby reducing the SNR. 
A notable instance is observed in the bottom plot ($\gamma=10$), where the intense wave input results in a negative SNR for at least 10 seconds of sampling before it can even be identified (0 dB).
\\
To that end, our proposed DL algorithm comes to the rescue, enabling a learning-based extraction of the rotation signal from the obscuring dynamics, accounting for varying intensities.
Fig.~\ref{fig:loss} depicts the gyrocompassing error through the model validation loss, comparing performance against the number of epochs and increasing $\gamma$ values. 
\\
Firstly, a noticeable correlation emerges between low $\gamma$ values and a low root mean squared error (RMSE), likely attributed to the heightened frequency content that corrupts the GT signal. Secondly, the presence of frequency contents introduce learning instability, rendering the generalization task more challenging as the GT pattern becomes less recognizable.
\begin{table}[t]
\centering
\caption{Analyzing gyrocompassing errors across benchmark models and varying intensities.}
\renewcommand{\arraystretch}{1.8}
\begin{tabular}{c c|c|c|c|c|c|}
\multicolumn{2}{c}{} & \multicolumn{5}{c}{\textbf{RMSE [deg]}} \\ \cline{3-7} 
 & & \ N/D \ & $\gamma=0.1$ & $\gamma=0.5$ & $\gamma=1$ & $\gamma=10$ \\ 
\cline{3-7}\addlinespace\cline{2-7} 
\multirow{5}{*}{\rotatebox[origin=c]{90}{\textbf{Model}}} & \multicolumn{1}{|c|}{Wavelet} & 4.83&7.23&9.72&14.18&31.59 \\ \cline{2-7}
    & \multicolumn{1}{|c|}{Wiener} & 4.90&7.77&10.44&14.76&35.64 \\ 
    \cline{2-7} 
    & \multicolumn{1}{|c|}{SG} & 4.56&6.33&8.52&12.91&28.96
 \\ 
    \cline{2-7} 
    & \multicolumn{1}{|c|}{FIR} & 4.87&7.43&9.97&13.64&35.64 \\ 
    \cline{2-7} 
    & \multicolumn{1}{|c|}{\textbf{Ours}} & \textbf{3.26} & \textbf{5.13} & \textbf{5.88} & \textbf{7.23} & \textbf{13.91} \\ \cline{2-7}
\end{tabular} \label{t:errors}
\end{table} 
\\
Table~\ref{t:errors} summarizes the data from these plots and provides a comparison with four state-of-the-art filters: wavelet denoising, the Wiener filter, the Savitzky-Golay (SG) filter, and the finite impulse response (FIR) filter \cite{chen2006new, schafer2011savitzky, saramaki19934}. 
\\
While they are regarded as reliable benchmarks for signal enhancement, their lower error bounds fall short compared to our model, which exhibits improvement rates ranging from $30\%$ to $60\%$, depending on $\gamma$ values.
\\
This underperformance is attributed to the inherent model-based limitations, which rely on simplistic cause-and-effect relationships. When complex dynamics are introduced, these approaches struggle to refine noisy inputs, rendering them unsuitable for operational platforms with minimal error tolerance.

\section{Conclusions} \label{s:conclusions}
In this paper, we address the challenging task of enabling gyrocompassing in an underwater environment where external ocean currents may interact with the UUV body. Through the design of a simplified vehicle structure and a dynamic motion model mimicking wave disturbances, we have successfully established an integrated pipeline capable of simulating realistic hovering scenarios. Exhibiting noise in both instrumentation and mechanical vibration aspects, the synthesized signals demonstrate low SNR, significantly complicating the gyrocompassing task. To overcome this challenge, we propose and train a dedicated learning model, which is then compared with conventional filtering techniques. The results indicate a substantial improvement, often in tens of percentages, as our model learns to selectively process meaningful information, effectively 'ignoring' additive dynamics. These findings underscore the potential of learning-based approaches to redefine error limits, making off-the-shelf gyroscopes competent for underwater gyrocompassing.

\section*{Acknowledgement}
D.E. is supported by the Maurice Hatter foundation and the Bloom School Institutional Excellence Scholarship for outstanding doctoral students at the University of Haifa.

\appendix 
\subsection{Second-Order Ordinary Differential Equation} \label{appendix:a}
Presented below are the simplified coefficient matrices of the ODE, utilized for solving the system dynamics.

\subsubsection{Mass-Inertia Matrix}
To begin with, let us assume a simplified solid cylinder, with mass $m$, radius $r$, length $l$, and centered about the ${z}$-axis such that its 3D inertia tensor is
\begin{align}
\boldsymbol{I}_0 = 
\begin{bmatrix}
\frac{1}{2}m r^2 & 0 & 0 \\
0 & \frac{1}{12}m (3r^2 + l^2) & 0 \\
0 & 0 & \frac{1}{12}m (3r^2 + l^2)
\end{bmatrix} \ .
\end{align}
Ideally, the mass-inertia matrix of an ideal 3D axisymmetric object, whose center of gravity coincides with its body-frame origin ($\boldsymbol{r}_W=\boldsymbol{0}_3$), degenerates into
\begin{align}
\boldsymbol{M} = \operatorname{diag} \{ \, m, \, m, \, m, \, I_{xx}, \, I_{yy}, \, I_{zz} \, \} \ .
\end{align}
But since mass is concentrated in the lower half $xy$ plane,
\begin{align}
\boldsymbol{r}_W = \begin{bmatrix}
0 & 0 & -z' \end{bmatrix}^{\TT} \ ,
\end{align}
along port/starboard and fore/aft symmetries, about the $xz$ and the $yx$ planes, respectively, turns $\boldsymbol{M}$ into 
\begin{align}
\boldsymbol{M} = \begin{bmatrix}
m & 0 & 0 & 0       & m z_G'& 0 \\ 
0 & m & 0 & -m z_G' & 0     & 0 \\ 
0 & 0 & m & 0       & 0     & 0 \\ 
0 & -m z_G' & 0 & I_{xx} & 0 & 0 \\ 
m z_G' & 0 & 0 & 0 & I_{yy} & 0 \\ 
0 & 0 & 0 & 0 & 0 & I_{zz}
\end{bmatrix} \ .
\end{align}

\subsubsection{Restoring forces and moments}
In a frictionless medium, an initially set-in-motion pendulum persists in oscillating indefinitely, thanks to mechanical energy conservation. However, real-world scenarios introduce non-conservative forces, including hydrodynamic drag, turbulence, and sea currents. These external factors gradually dissipate the initial energy input, ultimately leading the pendulum to return to equilibrium.
\\
Similarly, UUVs adhere to analogous principles, wherein their longitudinal and lateral planes respond to external torques by revolving around the $\phi$ and $\theta$ angles, tilting both gravity and buoyancy components. 
%
Being bottom-heavy, the center of gravity is below the body-fixed origin, $\boldsymbol{r}_W = \begin{bmatrix} 0 & 0 & z_W \end{bmatrix}^{\TT}$, while the center of buoyancy is above at $\boldsymbol{r}_B = \begin{bmatrix} 0 & 0 & z_B \end{bmatrix}^{\TT}$, resulting in the restoring force being
\begin{align} \label{eq:restoring}
\boldsymbol{g}(\boldsymbol{\eta}) = 
\begin{bmatrix} \vrule \\ \ \boldsymbol{g}_1 \quad  \\ \vertbar \vspace{1mm} \\ \hline \vertbar \\ \boldsymbol{g}_2 \\ \vertbar \end{bmatrix} = 
\begin{bmatrix}
( \boldsymbol{W} - \boldsymbol{B} ) \text{s}_\theta \\ 
-( \boldsymbol{W} - \boldsymbol{B} ) \text{c}_\theta \text{s}_\phi \\ 
-( \boldsymbol{W} - \boldsymbol{B} ) \text{c}_\theta \text{c}_\phi \\ 
(z_W \boldsymbol{W} - z_B \boldsymbol{B}) \text{c}_\theta \text{s}_\phi \\ 
(z_W \boldsymbol{W} - z_B \boldsymbol{B}) \text{s}_\theta \\ 
0
\end{bmatrix} 
\ .
\end{align} 
As observed, an upright vehicle ($\phi=\theta=0$) would experience all reactive moments as zeros, except for the downward gravity force. Upon perturbation, all reactive forces and moments (except the yawing moment) manifest as non-homogeneous solutions, causing both roll and pitch angles to successfully revert to their initial states, while the heading angle will consistently drift away. To preserve the linearity of the ODE, a small-angle approximation is applied to \eqref{eq:restoring} and substituted into \eqref{eq:6dof-full}, resulting in a second-order differential relation.

\subsection{Simulated Solutions} \label{appendix:b}
Generating practical solutions for the system involves assuming zero initial conditions and applying the small-angle approximation. This is followed by substituting \eqref{eq:substitution} into \eqref{eq:6dof-final} to transition to an inertial frame, resulting in 
\begin{align}
\boldsymbol{I}_0 \, \ddot{\boldsymbol{\eta}}_2  (t) + \boldsymbol{D}_2 \, \dot{\boldsymbol{\eta}}_2 (t)  + \textbf{g}_{2} \, \boldsymbol{\eta}_2  (t) = \boldsymbol{{\tau}}_2 (t) \ ,
\end{align}
where $\boldsymbol{W} - \boldsymbol{B}$ takes the matrix form of $\textbf{G}_{2} = \operatorname{diag} (\boldsymbol{g}_2)$.
\\
Given that all moment inputs $\boldsymbol{{\tau}}_2(t)$ are time-deterministic, the Laplace transform of the system is given by 
\begin{align}
s^2 \boldsymbol{I}_0 \, {\boldsymbol{\eta}}_2 (s) + s \boldsymbol{D}_2 \, {\boldsymbol{\eta}}_2 (s)  + \textbf{G}_2 \, \boldsymbol{\eta}_2  (s) = \boldsymbol{{\tau}}_2 (s) \ ,
\end{align}
resulting in the transfer function of the orientational response 
\begin{align} \label{eq:Laplace}
\boldsymbol{H} (s) = \frac{\boldsymbol{\eta}_2 (s)}{\boldsymbol{\tau}_2 (s)}
= \big( s^2 \boldsymbol{I}_0  + s \boldsymbol{D}_2  + \textbf{G}_2 \big)^{-1} \ .
\end{align}
In the frequency domain ($s=j\omega$), each of the angular states ($\phi, \theta, \psi$) assumes a standard oscillatory form, comprising a scaling factor multiplied by the response term
\begin{align} \label{eq:oscillatory}
\boldsymbol{\eta}_i (\omega) = \left( \frac{ \boldsymbol{\tau}_i (\omega) }{ \boldsymbol{I}_{0,ii} } \right) \cdot \frac{1}{ (\omega_0^2 - \omega^2) - (2 \omega \zeta ) j } \ , 
\end{align}
where $i \in \{ \text{x, y, z}\}$, the system frequency is $\omega_0 = \sqrt{\frac{G_{ii}}{I_{ij}}}$, and its damping ratio is $\zeta = \frac{D_{ij}}{2\sqrt{G_{ij} I_{ij}}}$. Note that both terms are functions of the inertia tensor, restoring forces, and torsional friction. 
Given the diverse nature of ocean currents, input waves $\boldsymbol{\tau}_2 (t)$ can be modeled as either an impulse function or a sinusoidal wave, leading to distinct dynamic responses. For instance, a pitching moment ($\tau_{2,x}$) characterized as a step function would produce an oscillatory roll angle, as follows
\begin{align}
\phi(t) = 1 - e^{-\zeta \omega_0 t} \frac{\sin \left( \sqrt{1 - \zeta^2} \omega_0 t + \phi_0 \right)}{\sin(\phi_0)}
\end{align}
where the initial phase angle is given by $\phi_0 = \cos^{-1} (\zeta)$.

\bibliographystyle{unsrt}
\footnotesize{\bibliography{Ref}}

\begin{thebibliography}{10}

\bibitem{titterton2004strapdown}
David Titterton and John~L Weston.
\newblock {\em Strapdown inertial navigation technology}, volume~17.
\newblock IET, 2004.

\bibitem{farrell2008aided}
Jay Farrell.
\newblock {\em {Aided navigation: GPS with high rate sensors}}.
\newblock McGraw-Hill, Inc., 2008.

\bibitem{britting2010inertial}
Kenneth~R Britting.
\newblock {\em Inertial navigation systems analysis}.
\newblock Artech House, 2010, 2010.

\bibitem{chatfield1997fundamentals}
Averil~Burton Chatfield.
\newblock {\em Fundamentals of high accuracy inertial navigation}, volume 174.
\newblock Aiaa, 1997.

\bibitem{park1995covariance}
Heung~Won Park, Jang~Gyu Lee, and Chan~Gook Park.
\newblock Covariance analysis of strapdown {INS} considering gyrocompass characteristics.
\newblock {\em IEEE Transactions on Aerospace and Electronic Systems}, 31(1):320--328, 1995.

\bibitem{renkoski2008effect}
Benjamin~Matthew Renkoski.
\newblock {\em The effect of carouseling on {MEMS-IMU} performance for gyrocompassing applications}.
\newblock PhD thesis, Massachusetts Institute of Technology, 2008.

\bibitem{klein2022data}
Itzik Klein.
\newblock Data-driven meets navigation: Concepts, models, and experimental validation.
\newblock In {\em 2022 DGON Inertial Sensors and Systems (ISS)}, pages 1--21. IEEE, 2022.

\bibitem{cohen2023inertial}
Nadav Cohen and Itzik Klein.
\newblock Inertial navigation meets deep learning: A survey of current trends and future directions.
\newblock {\em arXiv preprint arXiv:2307.00014}, 2023.

\bibitem{cohen2024kit}
Nadav Cohen and Itzik Klein.
\newblock {A-KIT}: Adaptive {Kalman}-informed transformer.
\newblock {\em arXiv preprint arXiv:2401.09987}, 2024.

\bibitem{groves2015principles}
Paul~D Groves.
\newblock Principles of {GNSS}, inertial, and multisensor integrated navigation systems, [book review].
\newblock {\em IEEE Aerospace and Electronic Systems Magazine}, 30(2):26--27, 2015.

\bibitem{woodman2007introduction}
Oliver~J Woodman.
\newblock An introduction to inertial navigation.
\newblock Technical report, University of Cambridge, Computer Laboratory, 2007.

\bibitem{engelsman2023parametric}
Daniel Engelsman, Yair Stolero, and Itzik Klein.
\newblock Parametric and state estimation of stationary {MEMS-IMUs}: A tutorial.
\newblock {\em arXiv preprint arXiv:2307.08571}, 2023.

\bibitem{nahon1996simplified}
Meyer Nahon.
\newblock A simplified dynamics model for autonomous underwater vehicles.
\newblock In {\em Proceedings of Symposium on Autonomous Underwater Vehicle Technology}, pages 373--379. IEEE, 1996.

\bibitem{prestero2001verification}
Timothy Timothy~Jason Prestero.
\newblock {\em Verification of a six-degree of freedom simulation model for the REMUS autonomous underwater vehicle}.
\newblock PhD thesis, Massachusetts institute of technology, 2001.

\bibitem{gomes2005underwater}
Sebasti{\~a}o C{\'\i}cero~Pinheiro Gomes, Carlos Eduardo~Motta Moraes, PLJ Drews~Jr, Tom{\'a}s~Garcia Moreira, and Adilson~Melcheque Tavares.
\newblock Underwater vehicle dynamic modeling.
\newblock In {\em 18th Int. Cong. of Mechanical Engineering-COBEM}, volume~5, 2005.

\bibitem{fossen1999guidance}
Thor~I Fossen.
\newblock Guidance and control of ocean vehicles.
\newblock {\em University of Trondheim, Norway, Printed by John Wiley \& Sons, Chichester, England, ISBN: 0 471 94113 1, Doctors Thesis}, 1999.

\bibitem{newman2018marine}
John~Nicholas Newman.
\newblock {\em Marine hydrodynamics}, volume 40th Anniversary Edition, pages 321--332.
\newblock The MIT press, 2018.

\bibitem{engelsman2023towards}
Daniel Engelsman and Itzik Klein.
\newblock Towards learning-based gyrocompassing.
\newblock {\em arXiv preprint arXiv:2312.12121}, 2023.

\bibitem{xue1997three}
Ming Xue.
\newblock {\em Three-dimensional fully-nonlinear simulations of waves and wave body interactions}.
\newblock PhD thesis, Massachusetts Institute of Technology, 1997.

\bibitem{chen2006new}
Jingdong Chen, Jacob Benesty, Yiteng Huang, and Simon Doclo.
\newblock New insights into the noise reduction {Wiener} filter.
\newblock {\em IEEE Transactions on audio, speech, and language processing}, 14(4):1218--1234, 2006.

\bibitem{schafer2011savitzky}
Ronald~W Schafer.
\newblock What is a {Savitzky-Golay} filter? [lecture notes].
\newblock {\em IEEE Signal processing magazine}, 28(4):111--117, 2011.

\bibitem{saramaki19934}
Tapio Saram{\"a}ki.
\newblock 4 finite impulse response filter.
\newblock {\em Handbook for digital signal processing}, page 155, 1993.

\end{thebibliography}
\end{document}